\documentclass[epj]{svjour}
\usepackage{graphics}
\begin{document}
\title{Proton radioactivity with a Yukawa effective interaction}
\author{T. R. Routray\inst{1}, S. K. Tripathy\inst{1}, B. B. Dash\inst{1}, B. Behera\inst{1} and D.N. Basu\inst{2} 
\thanks{ trr1@rediffmail.com; bibhutiphy@rediffmail.com; dnb@veccal.ernet.in }
}                     
%
%
\institute{P.G. Department of Physics, Sambalpur University, Jyoti Vihar, Burla, Orissa 768019, India \and Variable  Energy  Cyclotron  Centre, 1/AF Bidhan Nagar, Kolkata 700 064, India.}
\date {\today} 
\abstract{ 
    The half lives of proton radioactivity of proton emitters are investigated theoretically. Proton-nucleus interaction potentials are obtained by folding the densities of the daughter nuclei with a finite range effective nucleon-nucleon interaction having Yukawa form. The Wood-Saxon density distributions for the nuclei used in calculating the nuclear as well as the Coulomb interaction potentials are predictions of the interaction. The quantum mechanical tunneling probability is calculated within the WKB framework. These calculations provide reasonable estimates for the observed proton radioactivity lifetimes. The effects of neutron-proton effective mass splitting in neutron rich asymmetric matter as well as the nuclear matter incompressibility on the decay probability are investigated.   
} 

\authorrunning {T. R. Routray et al. }
\titlerunning {Proton radioactivity . .}
\maketitle
\noindent
\section{Introduction}
\label{section1}

    In recent years new half life measurements have been performed in order to have a better understanding of the proton and alpha decay processes in the region of proton-rich nuclei \cite{So02}. For these nuclei the Q value for proton emissions is positive and therefore there is a natural tendency to shed off excess protons. These data are very useful for the analysis of possible irregularities in the structure of these proton-rich nuclei \cite{Da96,Pa96}. They are also of great interest in rapid proton capture nucleosynthesis processes. Some new results for proton radioactivity in this region of proton-rich nuclei have indicated that the proton emission mode is rather competitive with the alpha decay process \cite{Pa96,En96,Le96}. Proton radioactivity may be used as a tool to obtain spectroscopic information because the decaying proton is unpaired in the orbit. These decay rates are sensitive to the Q values and the orbital angular momenta which in turn help to determine the orbital angular momenta of the emitted protons.  

    Since the observation of proton radioactivity is comparatively recent, several theoretical approaches that have been employed to study this exotic process, such as the distorted-wave Born approximation \cite{Ab97}, the density dependent M3Y (DDM3Y) effective interaction \cite{BCS05,MG07}, the effective interaction of Jeukenne, Lejeune and Mahaux (JLM) \cite{MG07}, the unified fission model \cite{Ba05}, the coupled-channels approach \cite{De06} and the effective/generalized liquid drop models \cite{Gu99,Do09} are also quite recent \cite{Ho05,La98}. In the present work, quantum mechanical tunneling probability is calculated within the WKB framework using proton-nucleus interaction potentials obtained from folding the density of the residual daughter nucleus with a finite range effective nucleon-nucleon interaction having  a single Yukawa term (YENI) in the finite range part \cite{Be02,Be05}. These calculations provide good estimates for the observed proton radioactivity lifetimes. In the present calculation we shall examine the effects pf neutron-proton (n-p) effective mass splitting as well as the nuclear matter incompressibility, $K(\rho_0)$, on the decay probability of the proton emitters. The n-p effective mass splitting is connected to the momentum dependence aspect of neutron and proton mean fields in asymmetric nuclear matter (ANM). Theoretical predictions of different models on this important issue can be divided into two distinct groups depending on whether the neutron effective mass goes above the proton one \cite{Zu99,Li04,Ma04,Zu05,Da05,Sa05} or the other way around \cite{Ho01,Ku97,Gr01,Li02,Ch97}. Experimental as well as theoretical attempts \cite{Ri04,Le06} to resolve the problem has not been successful as yet. In the work in Ref.\cite{Be05} it is shown that if the finite range exchange interaction acting between an unlike nucleon pair, $v_{ex}^{ul}(r)$, is stronger compared to the finite range exchange interaction, $v_{ex}^l$, between a pair of like nucleon, the neutron effective mass will be predicted to go over the proton one in neutron rich ANM. On the other hand if $v_{ex}^l(r)$ is stronger compared to $v_{ex}^{ul}(r)$ then proton effective mass will go over the neutron one. The results of Dirac-Brueckner-Hartree-Fock (DBHF) calculations \cite{Ma04,Da05} along with the decreasing trend of Lane potential extracted from experimental data on nucleon-nucleus scattering and reaction \cite{La62,Ho94,Ho72,Ko03} has led to accept amongst a larger community that neutron effective mass goes above the proton one in neutron rich matter although the controversy is not yet completely resolved. Under the circumstance the magnitude of effective mass splitting remains as an open problem as different models give widely divergent results. Similarly, the nuclear matter incompressibility is another important quantity whose value ranges between $240\pm20$ MeV as has been estimated from studies of isoscalar giant monopole resonances in nuclei. In the present work we shall also examine the possible effect of the variations of these nuclear matter parameters on the calculated proton half-lives of the proton emitters. 

    In section 2 the calculation of proton-nucleus (p-N) interaction potential for any general effective interaction has been discussed. The formalism has been extended to the calculation with the YENI. The determination of the parameters required in the study of the nuclear matter is briefly discussed. The fixation of the free parameter of the interaction along with the Wood-Saxon (WS) density distribution for the nuclei have been obtained by adopting a simultaneous minimization procedure to reproduce the binding energies of nuclei. Contributions of the different parts of YENI to the p-N nuclear potential and the self-consistent evaluation of the finite range exchange part are provided. In section 3 WKB tunneling procedure for calculation of decay probability of emitted proton has been discussed. The last section contains discussions of the results obtained and conclusions.

\noindent
\section{The proton-nucleus interaction potentials}
\label{section2}

\subsection{The folded proton-nucleus interaction potential}
 
    The proton-nucleus potential is obtained by folding the density distribution of the nucleus over the interaction of the incident proton with nucleons of the nucleus. It is given by \cite{Si73}

\begin{eqnarray}
 V_N(\vec{r}) =& \int [\rho_p (\vec{r'}) v_d^{pp}(|\vec{r} - \vec{r'}|) 
+ \rho_n (\vec{r'}) v_d^{pn}(|\vec{r} - \vec{r'}|) ] d^3r' \nonumber\\
& + \int \rho_p (\vec{r},\vec{r'}) j_0(k(\vec{R})| \vec{r} - \vec{r'}|)v_{ex}^{pp}(|\vec{r} - \vec{r'}|) d^3r' \nonumber\\
&+ \int \rho_n (\vec{r},\vec{r'}) j_0(k(\vec{R})| \vec{r} - \vec{r'}|)v_{ex}^{pn}(|\vec{r} - \vec{r'}|) d^3r' \nonumber\\
& +~~rearrangement~~terms
\label{seqn1}
\end{eqnarray}
\noindent
where $\vec{r}$ and $\vec{r'}$ are the distances of the incident proton and the nucleon of the nucleus, respectively, from the origin taken at the center of the nucleus. The last term in Eq.(1) is the rearrangement term that arises from the explicit density dependence of the effective interaction. $\rho_i (\vec{r},\vec{r'}), i=p,n$ are the density matrices that take non-local effects into account, and $v_{d/ex}$ is the direct/exchange part of the effective interaction averaged over space, spin and isospin of both the interacting nucleons. $j_0$ is the zeroth order spherical Bessel function. $k(\vec{R})$ is the wave number of the incident proton at the center of mass $\vec{R}$ of the incident proton and nucleon of the nucleus and is given by,

\begin{equation}
 k(\vec{R}) = \sqrt{\frac{2\mu}{\hbar^2}(E_{cm}-V_N(\vec{R})-V_c(\vec{R}))}
\label{seqn2}
\end{equation}
\noindent
where $E_{cm}$, $V_N(\vec{R})$ and $V_c(\vec{R})$ are the center of mass energy, p-N interaction potential and Coulomb potential at the center of mass $\vec{R}$, respectively. The center of mass and relative coordinates are given by $\vec{R}=(\vec{r} + \vec{r'})/2$ and $\vec{t}=(\vec{r} - \vec{r'})$ respectively. It may be seen from Eq.(1) that in the calculation of $V_N(\vec{r})$ the knowledge of $k(\vec{R})$ is required in which $V_N(\vec{R})$ appears and hence requires a self-consistent calculation. The total interaction energy between the proton and the residual daughter nucleus $V(\vec{r}) = V_N(\vec{r}) + V_C(\vec{r}) + \hbar^2 l(l+1) / (2\mu r^2)$, the sum of the nuclear interaction energy, the Coulomb interaction energy and the centrifugal barrier where $l$ is the angular momentum carried away by the proton-daughter nucleus system. Here $\mu = M_p M_d/M_A$  is the reduced mass, $M_p$, $M_d$ and $M_A$ are the masses of the proton, the daughter nucleus and the parent nucleus respectively, all measured in the units of MeV$/c^2$. 

\subsection{Simple finite range effective interaction and the proton-nucleus potential}

    The simple parameterization of finite range effective interaction \cite{Be05} used in this work for calculating proton radioactivity of the spontaneous proton emitters is given by,

\begin{eqnarray}
 v_{eff}(\vec{r} - \vec{r'}) = t_0(1+x_0 P_\sigma)\delta(\vec{r} - \vec{r'}) \nonumber \\
+\frac{t_3}{6}(1+x_3 P_\sigma) [ \frac{\rho(\vec{R})}{1+b\rho(\vec{R})} ]^\gamma \delta(\vec{r} - \vec{r'}) \nonumber \\
+(W + BP_\sigma - HP_\tau - MP_\sigma P_\tau)  f_\alpha (|\vec{r} - \vec{r'}|)
\label{seqn3}
\end{eqnarray}
\noindent
where $f_\alpha (|\vec{r} - \vec{r'}|)$, is a short range interaction of conventional form, such as, Yukawa, Gaussian or exponential and specified by a single parameter $\alpha$, the range of interaction. This effective interaction contains altogether eleven adjustable parameters, namely,  $t_0, x_0, t_3, x_3, b, \gamma, W, B, H, M$ and $\alpha$. $P_\sigma$=$(1+\vec{\sigma_1}.\vec{\sigma_2})/2$ and $P_\tau$=$(1+\vec{\tau_1}.\vec{\tau_2})/2$ are the spin and isospin exchange operators respectively. This interaction has been used in the studies of momentum and density dependence of both symmetric and asymmetric nuclear matter at zero and finite temperatures \cite{Be02,Be05} as well as in the calculation of bulk properties of neutron stars \cite{Be07} and equation of state (EOS) of beta stable $n+p+e+\mu$ matter, i.e., neutron star matter(NSM) \cite{Be09}. In these studies we require a total of nine parameter combinations, namely, $\alpha, b, \gamma, \varepsilon_0^l, \varepsilon_0^{ul},  \varepsilon_\gamma^l,  \varepsilon_\gamma^{ul}, \varepsilon_{ex}^l$ and $ \varepsilon_{ex}^{ul}$ out of the total eleven interaction parameters for the complete description of asymmetric nuclear matter and their relations to the interaction parameters are given in Ref. \cite{Be07}. Out of these nine parameters required for a complete description of ANM only six, namely $\alpha, b, \gamma, (\varepsilon_0^l+\varepsilon_0^{ul}), (\varepsilon_\gamma^l+  \varepsilon_\gamma^{ul})$ and $(\varepsilon_{ex}^l+\varepsilon_{ex}^{ul})$ are required to describe the EOS of symmetric nuclear matter (SNM). The careful adjustment of these six parameters so as to provide a correct momentum dependence of the mean field as well as density dependence of the EOS in SNM is discussed in the Refs.\cite{Be07,Be98}. The crucial advantage of the procedure adopted to constrain these six parameters in SNM is that the momentum dependence of the mean field can be varied with out changing the density dependence of the EOS of SNM and vice-versa is also true. The momentum dependence of the mean field in SNM is decided by the finite range exchange strength parameter $(\varepsilon_{ex}^l+\varepsilon_{ex}^{ul})$ and the range $\alpha$, whereas, the stiffness of the EOS is determined by the parameter $\gamma$ in the exponent. Under the consideration that the interaction between pairs of like $(l)$ and unlike $(ul)$  nucleons have same range but differ in strength, the study of ANM now, requires the correct splittings of the three parameters $(\varepsilon_0^l+\varepsilon_0^{ul})$, $(\varepsilon_\gamma^l+  \varepsilon_\gamma^{ul})$ and $(\varepsilon_{ex}^l+\varepsilon_{ex}^{ul})$ into two specific channels for interaction between pairs of like and unlike nucleons. In absence of adequate constraints, either experimental or theoretical, to decide the splitting of these three strength parameters the procedure that we have adopted in our study of ANM has been discussed in Refs.\cite{Be05,Be07}. The splitting of finite range exchange strength parameter $(\varepsilon_{ex}^l+\varepsilon_{ex}^{ul})$ into like and unlike channels decides the n-p effective mass splitting in ANM. The possible range of splitting into the like channel, i.e., $\varepsilon_{ex}^{l}$, can be from 0 to $(\varepsilon_{ex}^l+\varepsilon_{ex}^{ul})$ and accordingly the exchange strength in the unlike channel $\varepsilon_{ex}^{ul}$ is decided. For $\varepsilon_{ex}^{l}$ in between 0 and $(\varepsilon_{ex}^l+\varepsilon_{ex}^{ul})/2$ the neutron effective mass is predicted to lie over the proton one and for $\varepsilon_{ex}^{l}$ in the range $(\varepsilon_{ex}^l+\varepsilon_{ex}^{ul})/2$ and $(\varepsilon_{ex}^l+\varepsilon_{ex}^{ul})$ the vice-versa is the case. For a given $\varepsilon_{ex}^{l}$ the splitting of one of the rest two strength parameters, $(\varepsilon_0^l+\varepsilon_0^{ul})$ and $(\varepsilon_\gamma^l+  \varepsilon_\gamma^{ul})$, can be decided by assuming a standard value of symmetry energy $E_s(\rho_0)$ at normal density $\rho_0$. The splitting of the remaining parameter is decided from the value of $E_s'(\rho_0)=\rho_0 \frac{dE_s(\rho_)}{d\rho}|_{\rho=\rho_0}$. In order to decide the value of $E_s'(\rho_0)$ we have assigned arbitrary values to it and calculated the EOS of NSM in each case solving the charge neutrality and beta stability conditions. It is found that for a characteristic value of $E_s'(\rho_0)$ the asymmetric contribution to the nucleonic part of the EOS of NSM (that solely determines the composition of normal neutron stars) gives stiffest behaviour that remains almost stationary within  a small range around this value of $E_s'(\rho_0)$ \cite{Be07}. This is referred as the universal high density behaviour of the asymmetric contribution of the nucleonic part of the EOS in NSM. We have considered this characteristic value of $E_s'(\rho_0)$ that corresponds to the stiffest behaviour. For the standard value of $E_s(\rho_0)=30$ MeV, the value of $E_s'(\rho_0)$ is obtained to be 21.51 MeV for the EOS having $\gamma=1/2$ and $\varepsilon_{ex}^{l}$= $(\varepsilon_{ex}^l+ \varepsilon_{ex}^{ul})/3$.  $E_s'(\rho_0)$ varies from 20.93 MeV to 22.08 MeV as $\varepsilon_{ex}^{l}$ changes from 0 to $(\varepsilon_{ex}^l+\varepsilon_{ex}^{ul})/2$ showing a small variation. Similarly $E_s'(\rho_0)$ also shows a slow variation on the choice of $\gamma$ where it varies from 20.98 MeV to 21.70 MeV as $\gamma$ varies from 1/3 to 2/3 (corresponding to nuclear matter incompressibility in the range 220 to 253 MeV) for $\varepsilon_{ex}^{l}$ =$(\varepsilon_{ex}^l+\varepsilon_{ex}^{ul})/3$. The slope parameter $L = 3 E_s'(\rho_0)$ thus predicted for the EOSs considered lye within the range obtained from simultaneous analysis of neutron skin thickness results in nuclei and neutron star properties \cite{Ch05,Li08,Ce09}. Now, with the knowledge of all these nine parameters we are still left with two interaction parameters free for the calculation of finite nucleus. Here we considered $t_0$ and $x_0$ of our interaction in Eq.(3) as the free parameters. We determine the parameter $t_0$ by using a simultaneous minimization along with the WS density distribution parameters to fit to the binding energy of $^{40}Ca$ nucleus. 

\noindent
\subsection{ Determination of the parameter $t_0$ and Wood-Saxon density distribution of nucleus }
\label{subsection1}

The total energy of a nucleus is given by $E=E^{nucl}+E^{Coul}+E^{CM}$, where $E^{nucl}$, $E^{Coul}$ and $E^{CM}$ are the contributions from nuclear, Coulomb and center-of-mass correction, respectively. The nuclear part of the energy for an effective interaction is given by,

\begin{eqnarray}
E^{nucl}=\frac{\hbar^2}{2M} \int [\tau_n (\vec{r}) + \tau_n (\vec{r}) ] d^3r \nonumber \\
+\sum_{s,s'=n,p}[\frac{1}{2} \int\int \rho_s(\vec{r}) \rho_{s'}(\vec{r'})v_d^{ss'}(|\vec{r}-\vec{r'}|)d^3r d^3r' \nonumber \\
+\frac{1}{2}\int\int \rho_s(\vec{r},\vec{r'})\rho_{s'}(\vec{r},\vec{r'}) v_{ex}^{ss'} (|\vec{r}-\vec{r'}|)d^3r d^3r'], 
\label{seqn4}
\end{eqnarray}
\noindent   
where, the first term is the kinetic energy, second and third terms are direct and exchange contributions of the nuclear interaction; $\tau_n (\vec{r}), \rho_s (\vec{r})$ and $\rho_s (\vec{r},\vec{r'})$ with $s=n,p$ are the respective kinetic energy densities, densities and density matrices which are expressed in terms of single particle wave functions as,

\begin{eqnarray}
\tau (\vec{r})=\Sigma_{i=1}^A \nabla\phi_i^*(\vec{r}).\nabla\phi_i(\vec{r}) \nonumber\\
\rho (\vec{r})=\Sigma_{i=1}^A \phi^*_i(\vec{r})\phi_i(\vec{r}) \nonumber\\
\rho (\vec{r},\vec{r'})=\Sigma_{i=1}^A \phi_i^*(\vec{r})\phi_i(\vec{r'})
\label{seqn5}
\end{eqnarray}
\noindent   
In these expressions $\phi_i(\vec{r})$ are single particle wave functions, where the subscript $i$ denotes all the quantum numbers. Instead of going into the Hartree-Fock calculation of single particle states we have adopted a theoretically transparent and numerically simplified approach where the density matrix expansion (DME) of Negele and Vautherin \cite{Ne72} is used for the density matrices in the exchange interaction term of the energy expression in Eq.(4). The density matrix under the DME can be expressed as,

\begin{eqnarray}
\rho (\vec{R}+\frac{\vec{t}}{2},\vec{R'}-\frac{\vec{t}}{2}) = \frac{3j_1(k_f(\vec{R})\vec{t})}{k_f(\vec{R})\vec{t}}\rho(\vec{R}) \nonumber \\
+ \frac{35j_3(k_f(\vec{R})\vec{t})}{2 k_f^3(\vec{R})}[\frac{1}{4}\nabla^2\rho(\vec{R})-\tau(\vec{R}) \nonumber \\
+\frac{3}{5} k_f^2(\vec{R})\rho(\vec{R})]+ . . .  ,
\label{seqn6}
\end{eqnarray}
\noindent
where, $j_1$ and $j_3$ are spherical Bessel functions of order 1 and 3 respectively, $k_f(\vec{R})$ is the Fermi momentum corresponding to density $\rho(\vec{R})$ at the center-of mass $\vec{R}$ of the two interacting nucleons. Now choosing the Fermi momentum in the form \cite{Ho98},

\begin{equation}
q_f^2(\vec{R})=\frac{5[\tau(\vec{R})-\frac{1}{4}\nabla^2\rho(\vec{R})]}{3\rho(\vec{R})},
\label{seqn7}
\end{equation}
\noindent
reduces the DME in eq.(6) to the well known Slater approximation $\rho (\vec{R} + \frac{\vec{t}}{2}, \vec{R'}-\frac{\vec{t}}{2}) = \frac{3j_1(q_f(\vec{R})\vec{t})}{q_f(\vec{R})\vec{t}}\rho(\vec{R})$ of the exchange term but with a modified Fermi momentum that accounts for the surface corrections up to second order in the Thomas-Fermi model. We can now express the nuclear part of the energy in eq.(4) as 

\begin{equation}
E^{nucl}=\int H(\vec{R})d^3R,
\label{seqn8}
\end{equation}
\noindent
where, $H(\vec{R})$ is the energy density given by,

\begin{eqnarray}
&&H(\vec{R}) = \frac{\hbar^2}{2M} [\tau_n (\vec{R}) + \tau_n (\vec{R})] \nonumber \\ 
&&\times \sum_{s,s'=n,p} \frac{1}{2}\Big[ \rho_s(\vec{r}) \int\rho_{s'}(\vec{r'}) v_d^{ss'}(|\vec{r}-\vec{r'}|) d^3r' \nonumber \\ 
&&+\rho_s(\vec{r})\rho_{s'}(\vec{r'}) \int \frac{3j_1(q_s(\vec{R})t)}{q_s(\vec{R})t} \frac{3j_1(q_{s'}(\vec{R})t)}{q_{s'} (\vec{R})t}v_{ex}^{ss'}(t)d^3t \Big]
\label{seqn9}
\end{eqnarray}
\noindent
with $t$ being the relative coordinate and $q_s$, $s=n,p$ is the corresponding modified Fermi momentum that can be defined from eq.(7). The calculation of energy of a nucleus now requires the knowledge of density and kinetic energy density. Wood-Saxon density distribution,

\begin{equation}
 \rho(\vec{r})=\frac{\rho_0}{1+exp[(r-c)/a]}
\label{seqn10}
\end{equation}
\noindent
 is taken for simplicity that gives good description in the intermediate and heavy mass region. Further we make the approximation that $\rho_n$ and $\rho_p$ are proportional to neutron and proton numbers $N$ and $Z$ respectively. The kinetic energy density $\tau_s(\vec{R})$, $s=n,p$ is taken to be the Thomas-Fermi one along with the second order correction, 

\begin{equation}
\tau_s(\vec{R}) = \frac{3}{5}k_s^2\rho(\vec{R}) +\frac{1}{36}\frac{[{\nabla\rho_s(\vec{R})}]^2}{\rho_s(\vec{R})}  
+ \frac{1}{3} \nabla^2\rho_s(\vec{R}), 
\label{seqn11}
\end{equation}
\noindent
where, $k_{n(p)}(\vec{R}) = [3\pi^2\rho_{n(p)}(\vec{R})]^\frac{1}{3}$ is the neutron (proton) Fermi momentum at density $\rho_{n(p)}(\vec{R})$. The WS parameters $\rho_0$, $c$ and $a$ are determined by minimizing the total energy, including Coulomb and center-of-correction, with respect to these parameters. The Coulomb energy of the nucleus, both direct and exchange parts, has been calculated for the WS charge distribution. In minimizing the total energy we have taken $a=0.47$ fm and used the normalization $A=\int \rho(\vec{r})d^3r$ to express $\rho_0$ in terms of $c$ that reduces the minimization of the total energy with respect to $c$ only. Now, varying $t_0$ the minimization with respect to $c$ is done for the experimental value of the total energy of $^{40}Ca$. The parameter $t_0=481.86$ MeV fm$^3$ thus obtained for the EOS having $\gamma=1/2$ and $(\varepsilon_{ex}^l+\varepsilon_{ex}^{ul})/3$ predicts the charge radius of $^{40}Ca$ to be 3.49 fm as well as the binding energies and charge radii of the closed shell nuclei over the periodic table to an satisfactory extent as given in Table-1. On varying $x_0$ from -1 to +1 the binding energies of $N \neq Z$ closed shell nuclei show a slow variation from a relatively smaller value to a higher value of the binding energies as compared to the results for $x_0=0$ given in table-1. In view of the small variation we have taken $x_0=0$ in rest of our calculations that makes the $t_0$ part of the interaction spin independent. With the value of the $t_0$ thus obtained for the EOS, the binding energies of the proton radioactive nuclei have been calculated using the same minimization procedure with respect to the WS parameters and it has been found that the results are reproduced within 1/2$\%$ of the experimental values for these nuclei. The WS density distributions thus obtained for these proton radioactive nuclei are used in the evaluation of the p-N interaction potentials in order to calculate the half-lives. 

\begin{table}[h]
\caption{ Binding energies $B$ and charge radii $r_c$ of the closed shell nuclei. The experimental values are given besides the calculated ones inside the parenthesis. }
\center
\begin{tabular}{ccc}
\hline
\hline
Nucleus & $B$ [MeV] & $r_c$ [fm] \\ 
 
\hline
$^{48}$Ca&415.403 (415.991)&3.681 (3.484) \\ \hline

$^{90}$Zr&780.928 (783.893)&4.373 (4.272) \\ \hline

$^{132}$Sn&1091.70 (1102.86)&4.936 (- - - -) \\ \hline

$^{208}$Pb&1628.246 (1636.446)&5.664 (5.505) \\ \hline

\hline
\end{tabular} 
\end{table}

\noindent
\subsection{ Nuclear part of p-N interaction potential with the YENI }
\label{subsection2}

The p-N nuclear interaction potential given in Eq.(1) for the YENI in Eq.(3) becomes,
                            
\begin{equation}
 V_N(\vec{r}) = V^{zero}_N(\vec{r}) + V^{finite}_{N,dir}(\vec{r}) + V^{finite}_{N,ex}(\vec{r}) + V_N^{rearr}(\vec{r}),
\label{seqn12}
\end{equation}
\noindent 
where $ V^{zero}_N$ contains both direct and exchange contributions from the zero range parts of the interaction, $V^{finite}_{N,dir(ex)}$ denotes the contribution from the finite range direct (exchange) part of the interaction and $V_N^{rearr}$ is the rearrangement contribution. These various contributions are given by

\begin{eqnarray}
&V^{zero}_N(\vec{r})  = \frac{t_0}{2}[(1-x_0)\rho_p(\vec{r})+(2+x_0)\rho_n(\vec{r})]  \nonumber\\ 
&+ \frac{t_3}{12} [(1-x_3)\rho_p(\vec{r})+(2+x_3)\rho_n(\vec{r})] (\frac{\rho(\vec{r})}{1+b\rho(\vec{r})})^\gamma \nonumber\\ 
&V^{finite}_{N,dir}(\vec{r}) =\frac{4\pi(W+B/2-H-M/2)}{\mu^2}[\frac{e^{-\mu r}}{r} \nonumber\\
&\int_0^r r'\rho_p(\vec{r'}) sinh(\mu r')dr'   
+\frac{sinh(\mu r) }{r} \int_r^\infty r'\rho_p(\vec{r'}) e^{-\mu r'}dr' ] \nonumber\\
&+\frac{4\pi(W+B/2)}{\mu^2} [\frac{e^{-\mu r}}{r} \int_0^r r'\rho_n(\vec{r'}) sinh(\mu r')dr'   \nonumber\\ &+\frac{sinh(\mu r) }{r}\int_r^\infty r'\rho_n(\vec{r'}) e^{-\mu r'}dr' ] \nonumber\\ 
&V^{finite}_{N,ex}(\vec{r}) =\frac{2\pi(M-W/2+H/2-B)}{\mu r}  \nonumber\\ 
&\times \int_0^\infty r' dr' \int_{|\vec{r}-\vec{r'}|}^{|\vec{r}+\vec{r'}|} \rho_p(\vec{R}) \frac{3j_1[q_p(\vec{R})t]} 
{q_p(\vec{R})t} j_0(k(\vec{R})t) e^{-\mu t}dt \nonumber\\ 
&+\frac{2\pi(M+H/2)}{\mu r}  \nonumber\\ 
&\times \int_0^\infty r' dr' \int_{|\vec{r}-\vec{r'}|}^{|\vec{r}+\vec{r'}|} \rho_n(\vec{R}) \frac{3j_1[q_n(\vec{R})t]} 
{q_n(\vec{R})t} j_0(k(\vec{R})t) e^{-\mu t}dt \nonumber\\ 
&V_N^{rearr}(\vec{r})= \frac{t_3}{12}[(1-x_3)\frac{\rho_n^2(\vec{r}) +\rho_p^2(\vec{r})}{2} +(2+x_3)\rho_n(\vec{r})\rho_p(\vec{r})] \nonumber\\
&\times \frac{\gamma \rho^{\gamma-1}(\vec{r})}{[1+b\rho(\vec{r})]^{\gamma+1}}.
\label{seqn13}
\end{eqnarray}
\noindent
In obtaining $V^{finite}_{N,ex}$ in Eq.(13), we have approximated the density matrices $\rho_i (\vec{r},\vec{r'}),~i=n,p$ by their respective Slater terms,

\begin{equation}
 \rho_i (\vec{r},\vec{r'}) \approx \frac{3j_1(q_i(\vec{R})t)}{q_i(\vec{R})t}\rho_i (\vec{R})
\label{seqn14}
\end{equation}
\noindent                                                  
with modified Fermi momentum $q_{n(p)}(\vec{R}) = [3\pi^2\rho_{n(p)}(\vec{R})]^{1/3}$ that can be defined from Eq.(7). The zeroth order Bessel function $j_0(k(\vec{R})t)$ appearing in the expression of $V^{finite}_{N,ex}(\vec{R})$ is a function of the wave number $k(\vec{R})$ of the emitted proton that contains the potential $V_N(\vec{R})$ itself as can be seen from Eq.(2) and hence required to be evaluated self consistently. 

The p-N nuclear part of the potential, $V_N(\vec{r})$, in the cases of different proton radioactive nuclei are calculated from Eqs.(12,13) for a given EOS with the WS density distributions of the nuclei obtained from the minimization procedure discussed in the last sub-section. We have considered altogether five EOSs, two cases of different p-n effective mass splittings and three cases of nuclear matter incompressibility. In case of each of the five EOSs the parameter $t_0$ is obtained as discussed in the last subsection and the binding energies of the nuclei are verified to be reproduced within the same accuracy in each case. The two cases of effective mass splitting in nuclear matter at normal density $\rho_0$ corresponding to the values of $\varepsilon_{ex}^l$ =  $(\varepsilon_{ex}^l+\varepsilon_{ex}^{ul})/6$ and $(\varepsilon_{ex}^l+\varepsilon_{ex}^{ul})/2$ are shown as a function of asymmetry $\beta=\frac{\rho_n-\rho_p}{\rho_n+\rho_p}$ in Figure-1. The results of the calculations of $V_N(\vec{r})$ for different radioactive nuclei for these two cases of $\varepsilon_{ex}^l$ having a given $\gamma$ value show little difference. In view of this insensitivity of the p-N interaction potential to the n-p effective mass splitting we have considered a representative value, $\varepsilon_{ex}^l$ =  $(\varepsilon_{ex}^l+\varepsilon_{ex}^{ul})/3$, in our subsequent calculations of p-N potentials for the three different cases of $\gamma$, namely, $\gamma=1/3, 1/2$ and $2/3$, corresponding to the values of nuclear matter incompressibility 220, 240 and 253 MeV, respectively. The results of $V_N(\vec{r})$ for these three EOSs are shown in Figure-2 for the case of $^{113}Cs$. The difference in the results in these three cases are small having the characteristic behaviour of small extension of the tail in case of lower incompressibility. We shall examine the effect of these variations in $V_N(\vec{r})$ on the proton half-lives. 

\noindent
\subsection{ Coulomb part of p-N interaction potential }
\label{subsection3}

The direct and exchange parts of the Coulomb interaction potential of a proton with a nucleus having charge distribution $\rho_p(r)$ are given by 

\begin{equation}
V_C^{dir}(r) = 4\pi e^2[\frac{1}{r}\int_0^r r'^2\rho_p(r') dr' + \int_r^\infty r' \rho_p(r') dr']
\label{seqn15}
\end{equation}
\noindent 
and

\begin{equation}
V_C^{ex}(r) = -e^2(\frac{3}{\pi})^\frac{1}{3} \rho_p^\frac{1}{3}(r'),
\label{seqn16}
\end{equation}
\noindent 
respectively. The total Coulomb potential $V_C(r) = V_C^{dir}(r) + V_C^{ex}(r)$. The Coulomb potential between the emitted proton and the daughter nucleus is calculated from Eqs.(15) and (16) using the WS proton distribution of the daughter nucleus. 
                          
\begin{table*}[htbp]
\caption{Comparison between the measured and theoretically calculated half lives of proton emitters. The experimental $Q$ values, half lives and $l$ values are from Ref. \cite{So02}. The results of the present calculations using the YENI folded potentials are compared with the experimental values along with the results of DDM3Y \cite{BCS05} and GLDM \cite{Do09}. The turning points $R_2$=$R_a$ and $R_3$=$R_b$ are for YENI folded potentials for the case of $\gamma$=1/2 and  $\varepsilon_{ex}^l=(\varepsilon_{ex}^l+\varepsilon_{ex}^{ul})/3$. Experimental errors in $Q$ values \cite{So02} and corresponding errors in calculated half lives are inside parentheses. Asterisk symbol in the parent nucleus denotes isomeric state. }
\center
\begin{tabular}{cccccccccccc}
\hline
\hline
Parent & $l$ & $Q^{ex}$ &$R_2=R_a$ &$R_3=R_b$&Measured &YENI&$S^{expt}_p$&$S^{th}_p$&GLDM&DDM3Y    \\ 
$^A Z$& $\hbar$ & MeV &[fm]&[fm]&$log_{10}T(s)$&$log_{10}T(s)$&&&$log_{10}T(s)$ &$log_{10}T(s)$\\ 
\hline
$^{105}Sb$&2&0.491(15)&6.61&134.30&2.049$^{+0.058}_{-0.067}$&2.01(46)&0.914&0.999&1.831&1.90(45)\\ 
$^{109}I$&2&0.829(3)&6.69&83.29&-3.987$^{+0.020}_{-0.022}$&-4.20(4)&0.612 &-----&-----&-4.31(5)\\ 
$^{112}Cs$&2&0.824(7)&6.72&88.61&-3.301$^{+0.079}_{-0.097}$&-3.10(11)&1.589 &-----&-----&-3.21(11)\\ 
$^{113}Cs$&2&0.978(3)&6.78&73.45&-4.777$^{+0.018}_{-0.019}$&-5.51(4)&0.185 &-----&-----&-5.61(4)\\ 
$^{145}Tm$&5&1.753(10)&6.70&56.27&-5.409$^{+0.109}_{-0.146}$&-5.25(7)&1.442&0.580&-5.656&-5.28(7)\\ 
$^{147}Tm$&5&1.071(3)&6.73&88.65&0.591$^{+0.125}_{-0.175}$&0.85(4)&1.816&0.581&0.572&0.83(4)\\ 
$^{147}Tm^*$&2&1.139(5)&7.25&78.97&-3.444$^{+0.046}_{-0.051}$&-3.38(6)&1.159&0.953&-3.440&-3.46(6)\\ 
$^{150}Lu$&5&1.283(4)&6.77&78.23&-1.180$^{+0.055}_{-0.064}$&-0.72(4)&2.884&0.497&-1.309&-0.74(4)\\ 
$^{150}Lu^*$&2&1.317(15)&7.27&71.79&-4.523$^{+0.620}_{-0.301}$&-4.37(15)&1.422&0.859&-4.755&-4.46(15)\\ 
$^{151}Lu$&5&1.255(3)&6.79&78.41&-0.896$^{+0.011}_{-0.012}$&-0.80(4)&1.247&0.490&-1.017&-0.82(4)\\ 
$^{151}Lu^*$&2&1.332(10)&7.32&69.63&-4.796$^{+0.026}_{-0.027}$&-4.88(10)&0.824&0.858&-4.913&-4.96(10)\\
$^{155}Ta$&5&1.791(10)&6.88&57.83&-4.921$^{+0.125}_{-0.125}$&-4.79(7)&1.352&0.422&-2.410&-4.80(7)\\ 
$^{156}Ta$&2&1.028(5)&7.37&94.18&-0.620$^{+0.082}_{-0.101}$&-0.39(7)&1.698&0.761&-0.642&-0.47(8)\\ 
$^{156}Ta^*$&5&1.130(8)&6.86&90.30&0.949$^{+0.100}_{-0.129}$&1.52(10)&3.724&0.493&0.991&1.50(10)\\ 
$^{157}Ta$&0&0.947(7)&7.48&98.95&-0.523$^{+0.135}_{-0.198}$&-0.41(12)&1.297&0.797&-0.170&-0.51(12)\\
$^{160}Re$&2&1.284(6)&7.43&77.67&-3.046$^{+0.075}_{-0.056}$&-3.01(7)&1.086&0.507&-3.111&-3.08(7)\\ 
$^{161}Re$&0&1.214(6)&7.55&79.33&-3.432$^{+0.045}_{-0.049}$&-3.44(7)&0.982&0.892&-3.319&-3.53(7)\\ 
$^{161}Re^*$&5&1.338(7)&6.94&77.47&-0.488$^{+0.056}_{-0.065}$&-0.73(7)&0.528&0.290&-0.677&-0.75(8)\\ 
$^{164}Ir$&5&1.844(9)&7.02&59.97&-3.959$^{+0.190}_{-0.139}$&-4.06(6)&0.793&0.188&-4.214&-4.08(6)\\ 
$^{165}Ir^*$&5&1.733(7)&7.03&62.35&-3.469$^{+0.082}_{-0.100}$&-3.66(5)&0.644&0.187&-3.460&-3.67(5)\\
$^{166}Ir$&2&1.168(8)&7.49&87.51&-0.824$^{+0.166}_{-0.273}$&-1.12(10)&0.506&0.415&-1.099&-1.19(10)\\ 
$^{166}Ir^*$&5&1.340(8)&7.01&80.67&-0.076$^{+0.125}_{-0.176}$&0.07(9)&1.400&0.188&-0.025&0.06(9)\\ 
$^{167}Ir$&0&1.086(6)&7.61&91.08&-0.959$^{+0.024}_{-0.025}$&-1.26(8)&0.500&0.912&-1.074&-1.35(8)\\ 
$^{167}Ir^*$&5&1.261(7)&7.03&83.82&0.875$^{+0.098}_{-0.127}$&0.55(8)&0.473&0.183&0.858&0.54(8)\\ 
$^{171}Au$&0&1.469(17)&7.67&69.09&-4.770$^{+0.185}_{-0.151}$&-5.01(16)&0.575&0.848&-4.872&-5.10(16)\\ 
$^{171}Au^*$&5&1.718(6)&7.12&64.25&-2.654$^{+0.054}_{-0.060}$&-3.18(6)&0.298&0.087&-2.613&-3.19(5)\\ 
$^{177}Tl$&0&1.180(20)&7.72&88.25&-1.174$^{+0.191}_{-0.349}$&-1.36(26)&0.652&0.733&-1.049&-1.44(26)\\ 
$^{177}Tl^*$&5&1.986(10)&7.20&57.43&-3.347$^{+0.095}_{-0.122}$&-4.63(6)&0.052&0.022&-3.471&-4.64(6)\\ 
$^{185}Bi$&0&1.624(16)&7.84&65.71&-4.229$^{+0.068}_{-0.081}$&-5.44(14)&0.062&0.011&-3.392&-5.53(14)\\ \hline
\hline
\end{tabular} 
\end{table*}

\noindent
\section{ Proton radioactivity}
\label{section3}

    In the present work, the tunneling probability of the protons is calculated in the WKB framework. The WKB method has been found to be quite satisfactory for the $\alpha$ decay half life calculations and somewhat better than the S-matrix method \cite{Ma06}. The barrier penetrability $P$ in the improved WKB \cite{ke35} framework for any continuous (rounded) potential barrier is given by

\begin{equation}
 P = 1/ [1 + \exp(K)]
\label{seqn17}
\end{equation}
\noindent
and the decay constant by $\lambda=\nu P S_p$ where $S_p$ is the spectroscopic factor and the assault frequency $\nu$ is calculated from $E_v=\frac{1}{2}h\nu$, the zero point vibration energy. The half life is obtained from $T_{1/2}=\ln2/\lambda$. The decay half life $T_{1/2}$ of the parent nucleus $(A, Z)$ into a proton and a daughter $(A_d, Z_d)$ can, therefore, given by

\begin{equation}
 T_{1/2} = [(h \ln2) / (2 S_p E_v)] [1 + \exp(K)]
\label{seqn18}
\end{equation}
\noindent
where the action integral $K$ within the improved WKB approximation is given by 

\begin{equation}
 K = (2/\hbar) \int_{R_a}^{R_b} {[2\mu (V(r) - E_v - Q)]}^{1/2} dr
\label{seqn19}
\end{equation}
\noindent
with $R_a$ and $R_b$ being its 2$^{nd}$ and 3$^{rd}$ turning points determined from the equations 

\begin{equation}
 E(R_a)  = Q + E_v =  E(R_b)
\label{seqn20}
\end{equation}
\noindent
whose solutions provide three turning points. The proton oscillates between the first and the second turning points and tunnels through the barrier at $R_a$ and $R_b$. The zero point vibration energy $E_v$ is assumed to be proportional to $Q$ value of the spontaneous emission of protons. For the present calculations, the zero point vibration energies used here are the same as given by Eq.(5) of Ref. \cite{Po86} but extended to protons and the experimental $Q$ values \cite{So02} are used. The spectroscopic factor appearing in the denominator in Eq.(18) contribute a term $-log S_p$ to log $T_{1/2}$.

\begin{figure}[htbp]
\vspace{1.1cm}
\resizebox{0.44\textwidth}{!}{%
\includegraphics{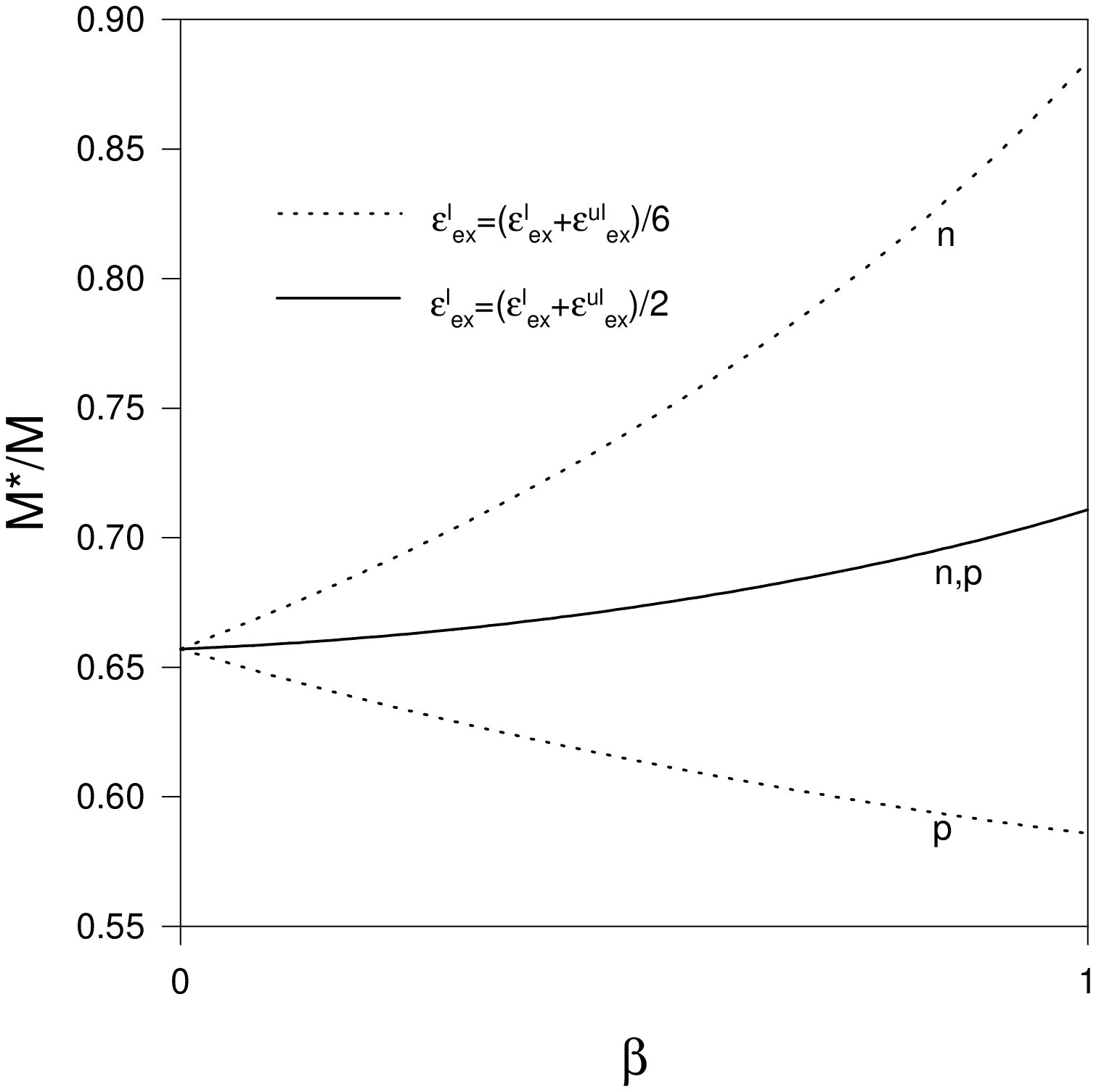}
}
\caption{ Neutron and proton effective masses $[M^*(k=k_{f_n,p},\rho_0,\alpha)/M]_n,p$ as a function of isospin asymmetry $\beta$ for the two cases of splittings of exchange strength parameter into like and unlike channels. For details see the text. }
\label{fig:1}       
\end{figure}
\noindent

\begin{figure}[htbp]
\vspace{2.95cm}
\resizebox{0.44\textwidth}{!}{%
\includegraphics{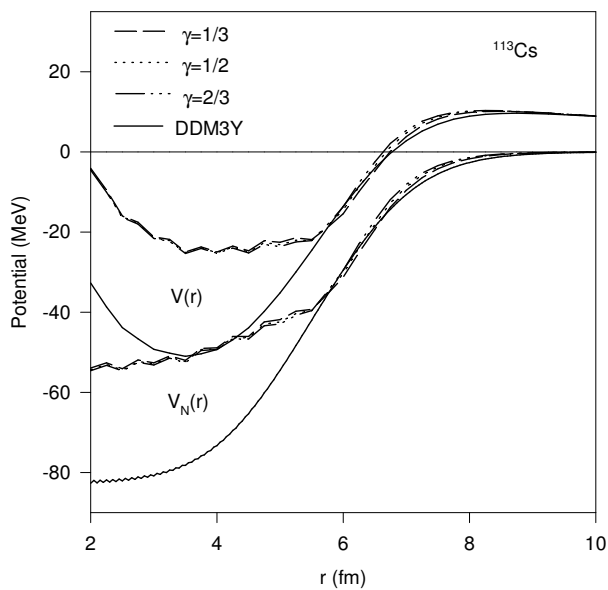}
}
\vspace{0.0cm}
\caption{ The p-N potential $V(r)$ and its nuclear part $V_N(r)$ are shown as a function of distance r in case of the nucleus $^{113}$Cs for DDM3Y and for three different cases of $\gamma=1/3,1/2$ and $2/3$. All three results corresponds to same $\varepsilon_{ex}^l=(\varepsilon_{ex}^l+\varepsilon_{ex}^{ul})/6$. For details see the text. }
\label{fig:2}       
\end{figure}
\noindent

\noindent
\section{Results and conclusion}
\label{section4}

    The half-lives in the cases of proton emitting nuclei away from proton drip line are calculated using the WKB barrier penetration method. The experimental $Q$ values together with their uncertainties are considered in calculating the penetration probabilities in different proton emitting nuclei. The nuclear part of the p-N interaction potential is calculated using the semiclassical approximation up to second order for the kinetic energy densities as well as for the density matrices. The DME used for the density matrices along with the modified Fermi momenta takes care of the surface corrections up to second order. The WS density distributions of the nuclei predicted by the interaction are used to calculate the nuclear and Coulomb parts of the p-N interaction potential. The direct part of the nuclear potential is evaluated exactly whereas the exchange part is approximate up to the second order correction of the density matrix expansion. The exchange part of the nuclear potential is evaluated self consistently. The interaction potential thus obtained for the YENI effective interaction in case of each nucleus is used to caculate the penetration probability. The half life is calculated from Eq.(15) under the consideration that the spectroscopic factor $S_p=1$. The results of different proton emitting nuclei are given in Table 2 for the EOS corresponding to $\gamma=1/2$ and $\varepsilon_{ex}^l = (\varepsilon_{ex}^l + \varepsilon_{ex}^{ul})/3$ along with the respective $Q$ and $l$ values. The results of other calculations using WKB barrier penetration are also listed in the same table for comparison together with the experimentally measured results. The agreement between the results of the present calculation and those of DDM3Y \cite{BCS08} are good being close to the experimental values compared to the JLM \cite{MG07} model. In the JLM model the WKB penetration probabilities are calculated from the interaction potentials by folding the JLM effective interaction with the densities of the nuclei obtained from the relativistic mean field model (RMF). For the cases of $^{147}Tm$, $^{150}Lu$, $^{156}Ta$, $^{156}Ta^*$, $^{177}Tl^*$ and $^{185}Bi$. The agreement of the calculated values in the present case as well as that of DDM3Y with the experimental results do not match well. These large deviations, particularly $^{177}Tl^*$ and $^{185}Bi$ could be brought down to reasonably close range of the experimental values in the GLDM model \cite{Do09} calculation by including the spectroscopic factors calculated from the RMF+BCS theory. The spectroscopic factor is found to be greatly affected by the proton shell structure and in turn contains shell effect to a large extent. The uncertianty in the present calculation of half lives attributed to the spectroscopic factor can be obtained from the relation \cite{Ab97}, $S^{expt}_p=\frac{T^{th}_{1/2}}{T^{ex}_{1/2}}$, where $T^{th(ex)}_{1/2}$ is the calculated (measured) haf life. The experimental spectroscopic factors $S^{expt}_p$ obtained from this relation in the present calculation is compared with the theoretical spectroscopic factors $S^{th}_p$ calculated using RMF+BCS model. The agreement qualitatively reproduces the general trend. It is worthwhile to mention here that the $S^{expt}_p$ values somewhat large compared to unity such as for the cases of $^{150}Lu$ and $^{156}Ta^*$, can be brought down from 2.884 and 3.724 to 2.317 and 2.350, respectively, if instead of mean values extrema values of measured and theoretical half lives are used. The discrepancies may be attributed to the fact that in the present calculations the shell and deformation effects are not considered rigorously. However, in the cases where the shell effects are not strong the calculated half lives matches with the experimental ones to a reasonable extent. 

    The effect of n-p effective mass splitting on the decay half lives has been examined by claculationg the half lives for the two cases of $\varepsilon_{ex}^l=(\varepsilon_{ex}^l+\varepsilon_{ex}^{ul})/3$ and $(\varepsilon_{ex}^l+\varepsilon_{ex}^{ul})/2$. In Fig.1, the neutron and proton effective masses $[M^*(k=k_{f_n,p},\rho_0,\alpha)/M]_n,p$ as a function of isospin asymmetry $\beta$ for the two cases of splittings of exchange strength parameter into like and unlike channels. The results for half lives are almost same as expected from the results of the interaction potentials in these two cases those differ within the line width. Thus the n-p effective mass splitting in finite nuclei on the proton decay has little effect and the reason may be attributed to the fact that the asymmetry as well as the Fermi momenta involved are small. On the other hand, the effect of the variation of nuclear matter incompressibility show observable effect on the decay half lives. With decrease in the value of $K(\rho_0)$, the decay half lives decreases. By decreasing $K(\rho_0)$ from 240 MeV (corresponding to $\gamma=1/2$) to 220 MeV ($\gamma=1/3$) the results of the calculated decay half lives decrease on the average by10$\%$. In case of $^{113}Cs$, the change in $log_{10}T(s)$ is from -5.55 (3.12 $\mu$sec) to 5.51 (2.83 $\mu$sec) as $\gamma$ decreases from 1/2 to 1/3. Similarly as $K(\rho_0)$ is increased from 240 MeV to 253 MeV by increasing $\gamma$ from 1/2 to 2/3, the calculated half lives increase on the average by 7.5 $\%$ and in $^{113}Cs$ $log_{10}T(s)$ is increased to -5.47 (3.39 $\mu$sec). In Fig.2, the nuclear interaction potentials $V_N(r)$ for these three EOSs of YENI have small differences in the tail region where the second turning point is located. Relatively slower rate of vanishing of the attractive nuclear potential $V_N(r)$ is observed in the tail region for the EOS corresponding to a lower value of $K(\rho_0)$ resulting in the shift of the potential barrier to higher distance. Accordingly the second turning point in this case will shift to a relatively higher distance compared to EOS corresponding to higher value of $K(\rho_0)$. In case of $^{113}Cs$ in Fig.2 the position of the second turning point has decreased from 6.84 fm to 6.72 fm as $K(\rho_0)$ has increased from 220 MeV to 253 MeV. This shift in the position of the second turning point is solely determined by the nuclear part of the interaction potential as the Coulomb and the centrifugal parts are same for all these EOSs. The position of the third turning point in all the cases is solely determined from the Coulomb interaction potential that gives the same result for the different EOSs considered. Thus the width of the potential barrier in case of EOS corresponding to a lower value of incompressibility decreases in comparison to the EOS having higher incompressibility resulting in the higher penetration probability. The values of our potential and that of DDM3Y \cite{BCS08} in the tail region in Fig.2 are close (more so when Coulomb and centrifugal potentials are added) and hence the agreement in the results for the half lives as given in Table.2. It shows that the predictions of the proton decay half lives in different models using the WKB penetration crucially depends on the nuclear potential in the tail region that determines the position of the second turning point and hence the penetration probability.   

Acknowledgement: This work is supported by the collaborative research scheme No. UGC-DAE-CSR-KC-CRS /2009/NP06/1354 of India. The work is covered under the SAP programme of School of Physics, Sambalpur University, India.

\end{document}